\documentclass{elsart3p}		

\usepackage{amssymb}
\usepackage[displaymath]{lineno}
\usepackage{graphicx}
\usepackage{subfigure}
\usepackage{captcont}

\journal{Astroparticle Physics}

\begin{document}

\begin{frontmatter}



\title{Measurement of sound speed vs. depth in South Pole ice for neutrino astronomy}


\author[Madison]{R.~Abbasi},
\author[Gent]{Y.~Abdou},
\author[Zeuthen]{M.~Ackermann},
\author[Christchurch]{J.~Adams},
\author[Madison]{J.~A.~Aguilar},
\author[Oxford]{M.~Ahlers},
\author[Madison]{K.~Andeen},
\author[Wuppertal]{J.~Auffenberg},
\author[Bartol]{X.~Bai},
\author[Madison]{M.~Baker},
% ...

\collab{(IceCube Collaboration)}

% ...
\address[Christchurch]{Dept.~of Physics and Astronomy, University of Canterbury, Private Bag 4800, Christchurch, New Zealand}
% ...
\address[Gent]{Dept.~of Subatomic and Radiation Physics, University of Gent, B-9000 Gent, Belgium}
% ...
\address[Madison]{Dept.~of Physics, University of Wisconsin, Madison, WI 53706, USA}
% ...
\address[Bartol]{Bartol Research Institute and Department of Physics and Astronomy, University of Delaware, Newark, DE 19716, USA}
\address[Oxford]{Dept.~of Physics, University of Oxford, 1 Keble Road, Oxford OX1 3NP, UK}
% ...
\address[Wuppertal]{Dept.~of Physics, University of Wuppertal, D-42119 Wuppertal, Germany}
\address[Zeuthen]{DESY, D-15735 Zeuthen, Germany}
\corauth[Vandenbroucke]{Corresponding author.  Address: Department of Physics, 366 LeConte Hall, University of California, Berkeley, CA 94720, USA (justinav@berkeley.edu).},
\thanks[Erlangen]{affiliated with Universit\"at Erlangen-N\"urnberg, Physikalisches Institut, D-91058, Erlangen, Germany}
\thanks[Bari]{on leave of absence from Universit\`a di Bari and Sezione INFN, Dipartimento di Fisica, I-70126, Bari, Italy}
\thanks[Kalmar]{affiliated with School of Pure and Applied Natural Sciences, Kalmar University, S-39182 Kalmar, Sweden}

\newpage

\begin{abstract}
We have measured the speed of both pressure waves and shear waves as a function of depth between 80 and 500~m depth in South Pole ice with better than 1\% precision.  The measurements were made using the South Pole Acoustic Test Setup ({SPATS}), an array of transmitters and sensors deployed in the ice at South Pole Station in order to measure the acoustic properties relevant to acoustic detection of astrophysical neutrinos.  The transmitters and sensors use piezoceramics operating at $\sim$5-25~kHz.  Between 200~m and 500~m depth, the measured profile is consistent with zero variation of the sound speed with depth, resulting in zero refraction, for both pressure and shear waves.  We also performed a complementary study featuring an explosive signal propagating from 50 to 2250~m depth, from which we determined a value for the pressure wave speed consistent with that determined with the sensors operating at shallower depths and higher frequencies.  These results have encouraging implications for neutrino astronomy: The negligible refraction of acoustic waves deeper than 200~m indicates that good neutrino direction and energy reconstruction, as well as separation from background events, could be achieved.
\end{abstract}


\end{frontmatter}



\begin{figure}[tbp]
\centering
\noindent\includegraphics[width=20pc]{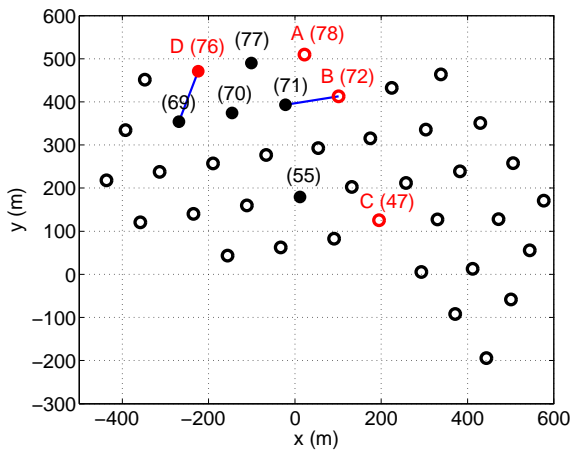}
\caption{Surface layout of the 40 strings constituting the IceCube array from February through November 2008.  The six holes in which the retrievable pinger was operated are indicated with filled circles.  The four holes with a SPATS string permanently deployed and frozen into the ice are indicated by SPATS ID letters.  IceCube hole ID numbers are given in parenthesis.  The two baselines used in this analysis are indicated by line segments.}
\label{geometry}
\end{figure}

\section{Introduction}

\begin{figure*}[t]
\centering
\noindent\includegraphics[width=40pc]{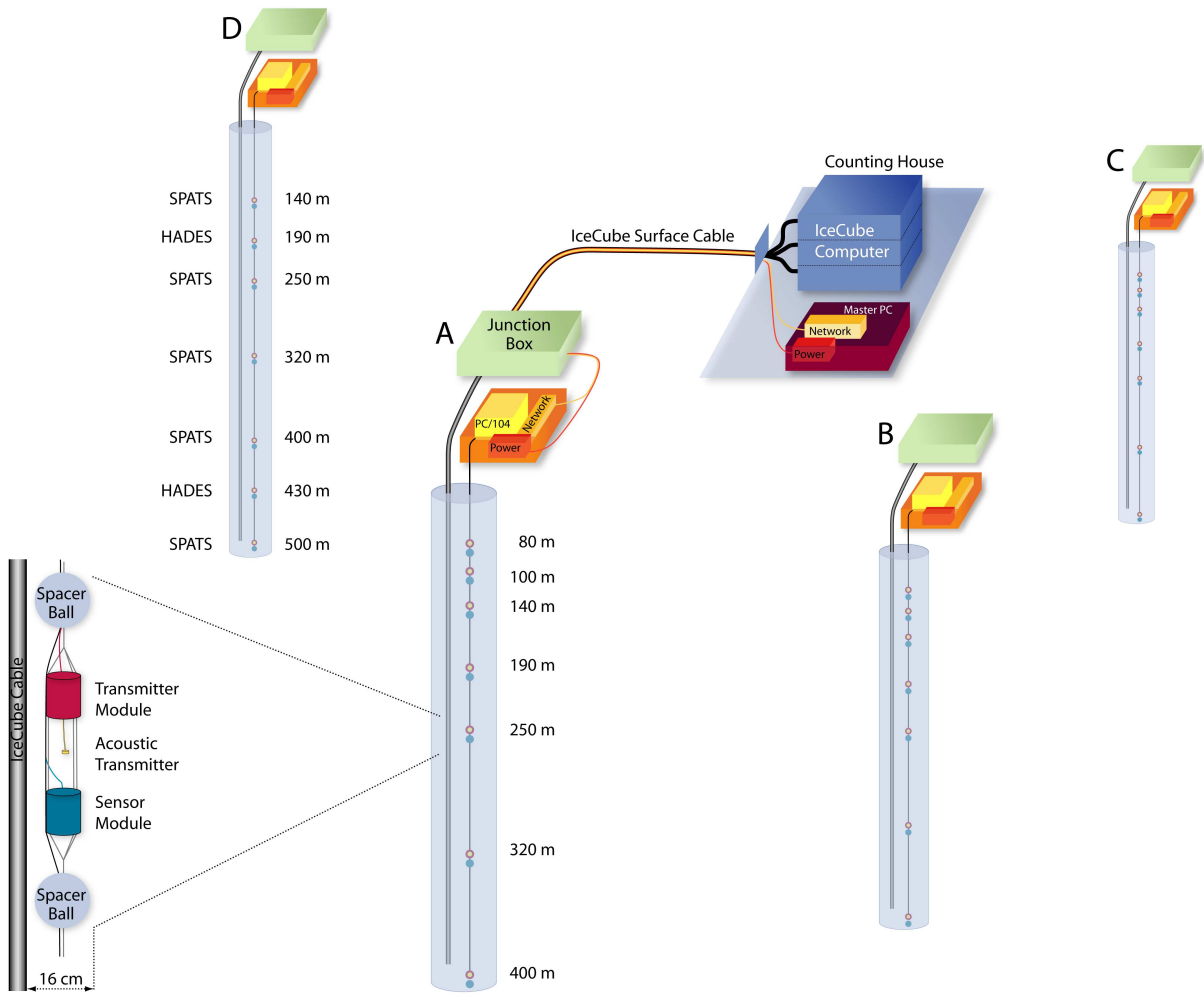}
\caption{Schematic of the SPATS array.}
\label{schematic}
\end{figure*}

South Pole ice is uniquely suited as a medium for detection of high-energy (10$^{11}$-10$^{20}$~eV) neutrinos of astrophysical origin.  The interactions of these neutrinos in ice produce optical, radio, and acoustic radiation, each of which therefore provides a possible method of detecting the neutrinos.  The optical method is well suited for neutrinos of energy up to 10$^{17}$~eV, while the radio and acoustic methods are well suited for neutrinos of higher energy.  Deep ice at the South Pole has been shown to be extremely transparent in optical wavelengths~\cite{Ackermann06}.  The AMANDA~\cite{Ackermann:2007km} and IceCube~\cite{Achterberg:2007bi} detectors have been developed to exploit this for optical neutrino detection.  Antarctic ice is even more transparent in radio wavelengths~\cite{Barwick05},~\cite{Besson:2007ek}, and the Radio Ice Cherenkov Experiment (RICE)~\cite{Kravchenko06} was operated to search for radio signals from astrophysical neutrinos.  \emph{In situ} measurements of the acoustic attenuation length in South Pole ice are underway~\cite{Vandenbroucke08}.

To detect the ``cosmogenic'' neutrinos of energy $\sim$10$^{17-19}$~eV produced by ultra-high-energy cosmic rays interacting with the cosmic microwave background radiation, a detector with effective volume on the order of 100~km$^3$ is necessary.  While the optical method is well understood and calibrated with atmospheric neutrinos, it is prohibitively expensive to scale to such a size.  The acoustic and radio methods, on the other hand, can in principle be used to instrument a large volume sparsely and achieve good sensitivity per cost in this energy range.

The acoustic radiation is produced by the ``thermoacoustic'' mechanism: A neutrino interacts to produce a shower of particles, which locally heats the medium, causing it to expand and produce a bipolar shock wave.  The pulse width (peak frequency) and shape depend on the sensor location relative to the shower.  The acoustic source is simply the region over which significant heat is deposited by the shower: a filament with length of a few meters and diameter of a few centimeters.  The filament is aligned along the incident neutrino direction.  The acoustic radiation pattern is a wide, flat disk perpendicular to the filament and therefore perpendicular to the neutrino direction~\cite{Learned79}.  The peak frequency is $\sim$30~kHz at a distance of 1~km from the source, for points near the center of the radiation pattern~\cite{Bevan07}.  

South Pole ice is predicted to be especially well suited for acoustic detection of extremely high-energy neutrinos~\cite{Price06}.  The neutrino-induced signal amplitude is larger in ice than in water due to its favorable elastic and thermal properties.  Furthermore, we have determined the background noise to be very stable in South Pole ice~\cite{Karg09}, in contrast to ocean water where it is highly variable on multiple time scales, resulting in the necessity of sophisticated trigger algorithms~\cite{Vandenbroucke05},~\cite{Kurahashi07}.

As a solid, ice also has the distinct advantage that it can support shear wave propagation.  If neutrinos produce shear waves in addition to pressure waves, a single acoustic sensor detecting both pulses could determine the distance to the interaction vertex as well as the particle shower energy.  Multiple sensors seeing some combination of pressure and shear waves could reconstruct the neutrino energy and direction better than if pressure waves alone were detected.  However, while much theoretical and experimental work has been done on pressure waves generated by the thermoacoustic mechanism, little work has been done on shear waves.  It has been argued on theoretical grounds~\cite{Boeser06} that shear wave production by the thermoacoustic mechanism is suppressed, but other mechanisms could produce shear waves and in any case laboratory measurements are necessary.

The speed of sound in ice has been studied in theory, in the laboratory, and in the field.  In addition to pure interest in elastic materials physics, the measurement has applications to both geophysics~\cite{martinb:ice-density},~\cite{Kohnen} and neutrino astronomy~\cite{Vandenbroucke08}.  At the South Pole, one measurement was made previously for pressure waves at seismic frequencies, for depths between 0~m and 186~m (i.e., in the layer of uncompactified surface snow, or ``firn''), using surface explosions~\cite{Weihaupt63}.

Beyond the South Pole, previous authors have also reported a variety of sound speed measurements in a wide range of conditions including laboratory and field measurements.  Field measurements have previously been made across the Antarctic and Greenland ice sheets and in temperate glaciers.  In principle the sound speed can vary from site to site due to differences in bubble concentration, temperature, and grain orientation.  The grain orientation as a function of position in a glacier is known as the ``fabric.''  The fabric can have a significant effect on the sound speed because the speed in monocrystalline ice varies by 7\% depending on the direction of propagation relative to the crystal axis~\cite{Price93}.  If the grain orientation is random, the sound speed is homogeneous and isotropic on macroscopic scales.  If there is non-random fabric, the sound speed can be inhomogeneous or anisotropic.

The South Pole Acoustic Test Setup (SPATS)~\cite{Boeser08} was installed to measure the acoustic properties of South Pole ice relevant to neutrino astronomy, in particular the sound speed profile, the background noise (both the noise floor and the impulsive transients), and the attenuation length.  Here we focus on the first of these: the sound speed as a function of depth.  Sufficiently mapping this profile \emph{in situ} allows precise reconstruction of the location of transient acoustic sources in the ice, which has now been achieved with SPATS~\cite{Vandenbroucke09} using the results presented here.

We report an \emph{in situ} measurement made using SPATS, which is comprised of transmitters and sensors deployed between 80 and 500~m depth and operating in the audible to ultrasonic band.  In addition to making an independent measurement of pressure wave speed in the firn and extending Weihaupt's measurements from the firn deep into the bulk ice, we have measured for the first time the speed of shear waves in both the firn and bulk ice.  Previously, the best estimate of South Pole shear wave speed was a model based on the pressure wave speed and Poisson's ratio~\cite{martinb:ice-density}, and only applied in the firn where pressure wave speed measurements were available.


\section{Experimental method}

\subsection{Frozen-in sensors}

The IceCube neutrino detector is currently under construction.  40 IceCube holes were drilled by the end of January 2008 (Figure~\ref{geometry}).  An additional 19 holes were drilled between December 2008 and January 2009.  Each IceCube hole contains a string with 60 digital optical modules between 1450~m and 2450~m depth.  Each hole is drilled with hot water to produce a standing water column $\sim$60~cm in diameter.  The instrumentation is then installed in the hole and the water column re-freezes around it.

The SPATS array consists of 4 strings, each deployed alongside an IceCube string in an IceCube hole.  A schematic of the array is given in Figure~\ref{schematic}.  Each SPATS string contains 7 acoustic ``stages.''  Each stage comprises one transmitter module and one sensor module.  For the measurement presented here, only the sensor of each stage was used and a separate, retrievable pinger was used instead of the frozen-in transmitters.  On Strings A, B, and C the stages are at 80, 100, 140, 190, 250, 320, and 400~m depth.  On String D the stages are at 140, 190, 250, 320, 400, 430, and 500~m depth.  Each acoustic sensor module contains 3 piezoelectric sensor channels (each with its own pre-amplifier), with the exception of the modules at 190~m and 430~m depth on String D.  Each of these two modules contains a single sensor channel of an alternative design (``Hydrophone for Acoustic Detection at South Pole'', HADES~\cite{Semburg09}).  The sensors are sensitive in the 5 to 100~kHz range.  The signal of the frozen-in transmitters (not used for this analysis) is broadband and peaked at $\sim$50~kHz.

The analog output of each channel is transmitted along copper cables to the surface, where it is digitized at 200 kilosamples per second by a rugged embedded computer (``String PC'') installed in a junction box buried 2~m beneath the snow surface.  Power, communications, and timing are distributed over surface cables several hundred meters long to each of the String PC's from an indoor server (``Master PC'') in the IceCube Laboratory.

In addition to digitizing the sensor waveforms, the String PC time stamps them.  Absolute time stamping is achieved with an IRIG-B signal routed to the String PC's from a GPS clock (Meinberg model \emph{GPS169PCI}) installed in the Master PC.  A single IRIG-B output from the clock is fanned out into four cables routed to the String PC's.  The GPS clock is specified to produce IRIG-B rising edges within $\pm$2~$\mu$s of absolute GPS time.  The delay introduced in the IRIG-B signals during propagation from Master PC to String PC is a few $\mu$s, negligible compared to other sources of timing uncertainty in this analysis.

To acquire the results presented here, each sensor channel was recorded continuously for 9~s in order to capture 9 consecutive pulses of the pinger which was operated at a 1~Hz repetition rate.  On each string the sensor channels were read out one by one and looped over sequentially, with 2~s of dead time between channels.  The strings recorded simultaneously, with one channel recording on each string at a time.  This scheme resulted in every channel of the array being recorded every 4~minutes.

\subsection{Retrievable pinger}

In addition to the permanently deployed array of sensors and transmitters, a retrievable pinger was operated in six water-filled IceCube holes, prior to IceCube string deployment in each hole, during the 2007-2008 season.\footnote{An upgraded version of the pinger was operated in four holes in the 2008-2009 season.  Here we focus on results from the 2007-2008 season.}  The pinger consisted of an isotropic piezoceramic emitter ball and a high voltage (HV) module.  The HV module consisted of a high voltage generator circuit contained in a steel pressure housing.  The emitter ball (model \emph{ITC-1001} from the International Transducer Company) produced a broadband pulse peaked in the $\sim$5-25~kHz range.  It was suspended $\sim$1.7~m below the HV module to reduce the effect of acoustic reflections off the steel housing.  The pinger was deployed on a steel-armored, four-conductor cable, which provided both the mechanical and electrical connection from the pinger to the surface.  It was lowered and raised from the surface with a winch.  The length of the cable was $\sim$2700~m, most of which remained spooled on the winch throughout the deployment.

On the surface, a GPS clock (Garmin model \emph{GPS 18 LVC}) was used to generate a 1 pulse per second (PPS) signal, with the rising edge of each pulse aligned to the start of each GPS second.  The PPS signal was routed over the armored cable to the pinger where it served as a trigger signal to pulse the pinger at 1~Hz with emission at known absolute times (modulo 1~s).  The rising edge of the PPS signal initiated charging of the HV pulser circuit, followed by discharge a time $t_e$ later, immediately resulting in acoustic emission.  This emission time delay introduced by the HV pulser was measured in the laboratory to be $t_e = $~1.9~$\pm$~0.05~ms over the range of temperatures in which the pinger operated ($-20$~$^\circ$C to $+20$~$^\circ$C).  The electrical pulse-to-pulse variation of the HV pulser module is negligible.

The electrical signal propagation speed through the $\sim$2700~m cable is 67\% of the speed of light in vacuum according to the manufacturer's specifications, resulting in a $\sim$13~$\mu$s predicted cable delay time.  This delay was verified in the laboratory to be on the order of 10~$\mu$s, negligible compared with other contributions to the timing uncertainty.  The GPS clock is specified to produce rising edges synchronized with absolute time within $\pm$1~$\mu$s.

Although the electrical pulse applied to the transmitter is monopolar, both the transmitter and the sensors ring.  This means that each pulse waveform contains many cycles (oscillations).  The rising edge of the first one is used to determine the acoustic signal propagation time.

The pinger pulsed at 1~Hz repetition rate while it was lowered from the surface to a maximum depth of 400-500~m and then raised back to the surface.  At each depth for which there was a frozen-in sensor on the recording SPATS strings, lowering was halted to keep the pinger stationary for 5 minutes.  This scheme guaranteed that every sensor channel of the SPATS array recorded one complete 9~s waveform while the pinger was stopped at each depth.

In addition to the expected pressure waves, shear waves were clearly detected for many pinger-sensor configurations.  Shear waves were previously detected from frozen-in SPATS transmitters, but it was a surprise to detect them from the pinger operating in water through which shear waves cannot propagate.  While the shear waves from the frozen-in transmitters were likely produced at the piezeoelectric transducers themselves, the shear waves from the pinger in water were likely produced by mode conversion at the water/ice interface (hole wall).  Such mode conversion would be suppressed if the incident angle were normal, but if the pinger was not in the center of the hole the incident angle was oblique and shear wave production was favored.  Pinger pressure wave and shear wave identification and characterization are presented in detail in~\cite{Vandenbroucke08}

In the 2008-2009 season the pinger was operated with a mechanical centralizer to keep it close to the center of the hole, and the shear wave production was suppressed in that data compared to the 2007-2008 data analyzed here.  The transmission (for both pressure and shear waves) and reflection (for pressure waves only) coefficients in terms of the incident angle are given by the Zoeppritz equations~\cite{Aki02}.  However, they are difficult to apply to this problem because the incident angle depends on the unknown lateral position of the pinger in the hole.

\begin{figure*}[tbp]	
\begin{center}
\subfigure[][]{
\label{waveform-a}
\noindent\includegraphics[width=18pc]{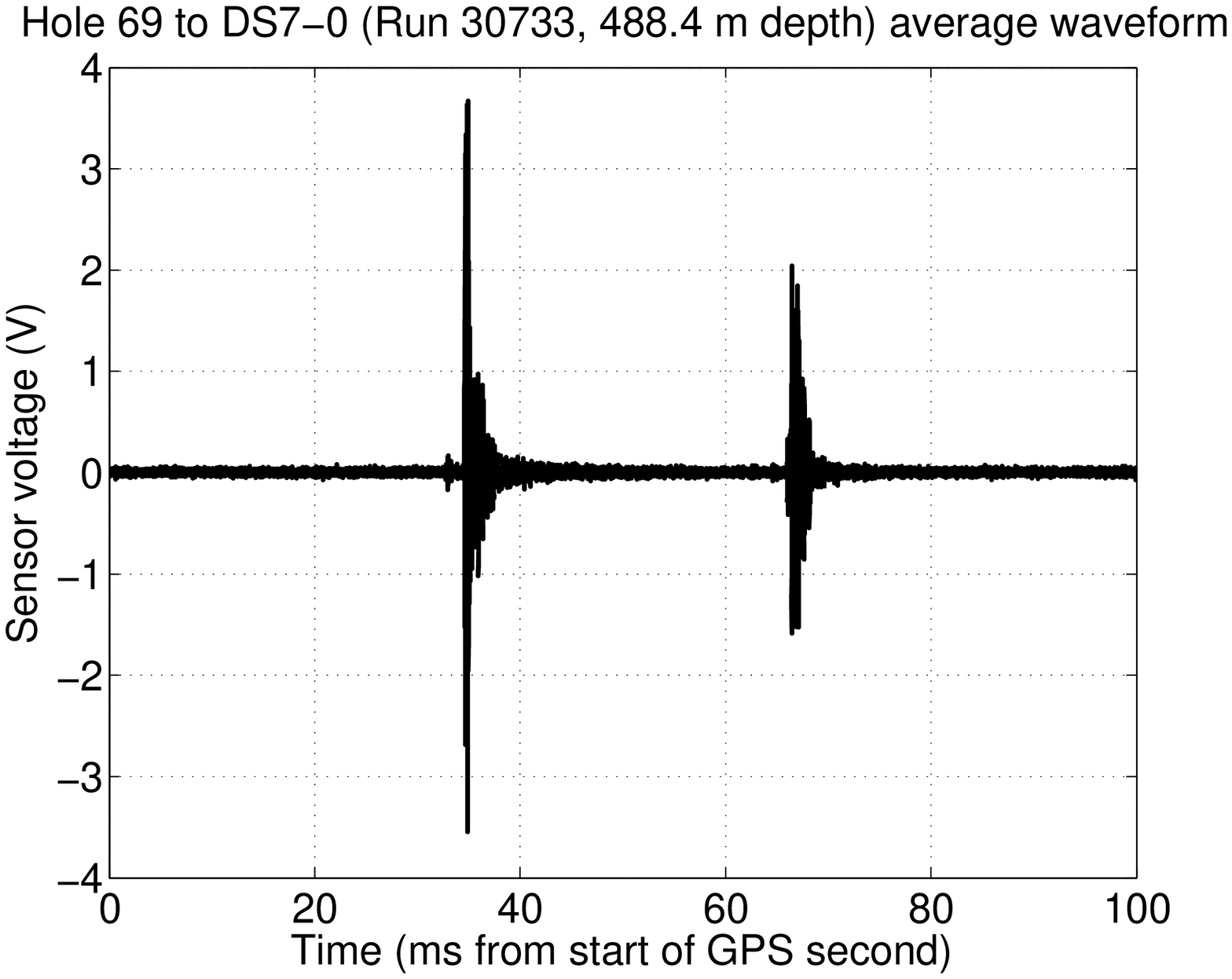}
}
\subfigure[][]{
\label{waveform-b}
\noindent\includegraphics[width=18pc]{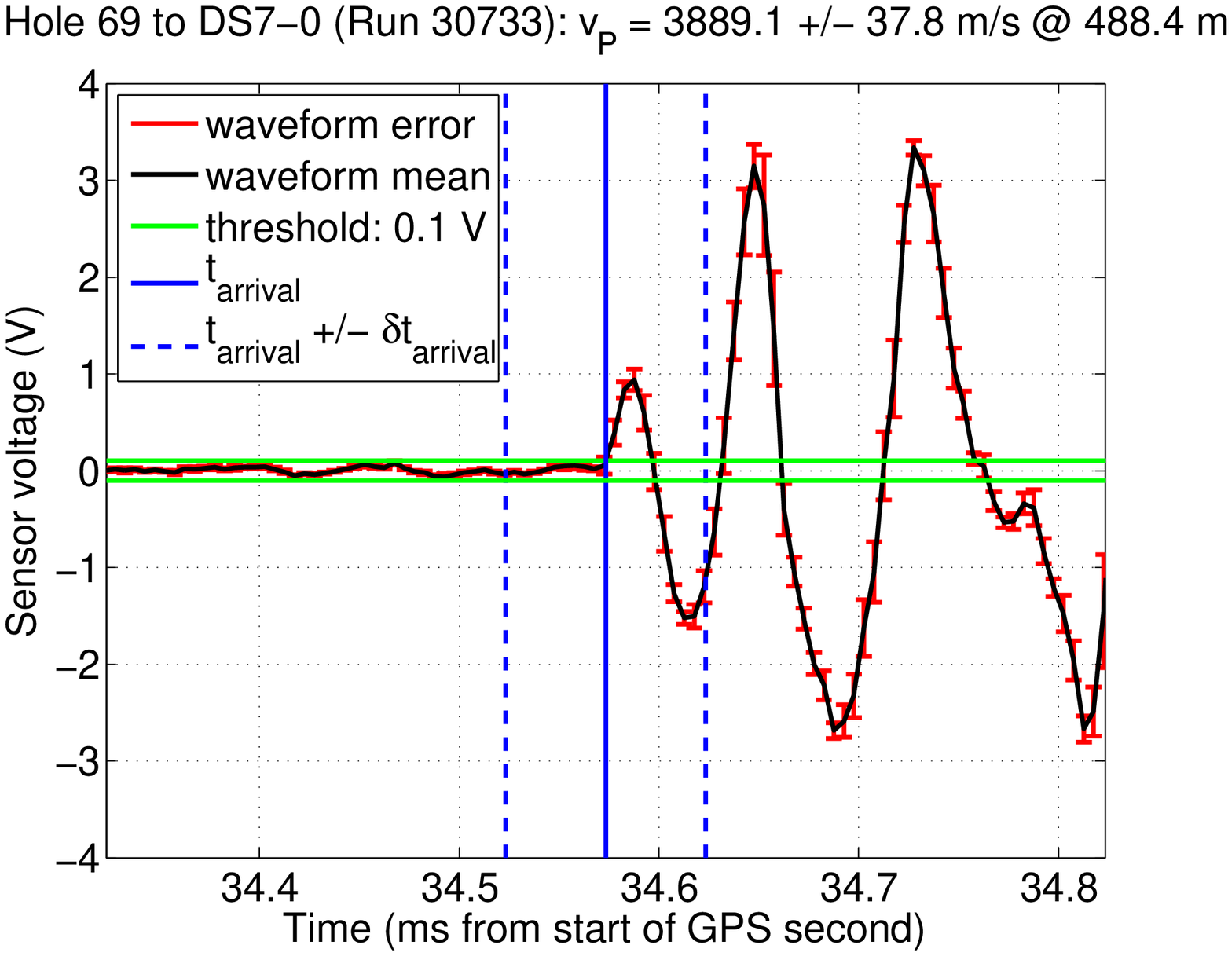}
}
\caption{An example waveform recorded by a sensor.  \subref{waveform-a} shows the full average waveform resulting from averaging 9 pulses, accounting for clock drift.  Both the pressure pulse and the shear pulse are clearly visible above the noise.  The small pre-pulse before the main pressure pulse is an acoustic pulse initiated by the rising edge of the pinger trigger signal (the main pulse is discharged by the falling edge of the trigger signal, 1.9~ms later).  \subref{waveform-b} shows a close-up of the beginning of the main pressure wave.  For each sample in the mean waveform, the uncertainty is estimated to be $\pm$1 standard error of the mean of the 9 samples contributing to the average.  The threshold used to determine the signal start time is shown, as are the signal start time and uncertainty of the start time.  Although the signal start time is clear in this example, for other waveforms it was unclear if the algorithm selected the correct first signal oscillation or was wrong by $\sim$one oscillation period.  An uncertainty of $\pm$0.05~ms was therefore assigned to the start time for all waveforms.}
\label{waveform}
\end{center}
\end{figure*}

\section{Data analysis}

\subsection{Geometry}
We analyzed two pinger-to-sensor hole combinations: Hole 69 to String D and Hole 71 to String B.  Two combinations were used both as a cross-check and to increase the number of depths included in the analysis.  The two hole combinations used for this analysis are nearest neighbors in the IceCube grid (125~m nominal spacing).  The horizontal distances between the holes (measured by a surveyor) are 124.9~m for Hole 69 to String D, and 124.6~m for Hole 71 to String B.

There are three contributions to the uncertainty in the horizontal separation between the pinger and sensor.  First, the center of each hole at the surface is determined by surveying to $\pm$0.1~m precision in each of the $x$ and $y$ coordinates.  Second, each IceCube hole has a radius of $\sim$0.3 m.  Assuming the pinger could be located laterally anywhere in the cylinder with equal probability, each of the $x$ and $y$ coordinates is within $\pm$0.17~m of the hole center at 68\% confidence.  Similar logic applies for the sensor.  Third, the drill head drifts laterally during drilling of each hole.  Using inclinometers located on the drill head, we estimate this drift to be $\pm$0.5~m in each coordinate.  This effect dominates the first two effects.  Therefore the uncertainty in the horizontal location of each of the pinger and sensor is $\pm$0.5~m and the uncertainty in the horizontal distance between the two is $\pm$0.5~m~$\times$~$\sqrt{2}$~=~$\pm$0.71~m.

The depth of each frozen-in SPATS sensor was verified during string deployment with a pressure sensor.  Each SPATS sensor depth is within $\pm$2~m of nominal.  The pinger depth was monitored in two ways: using pressure sensors and counting  the number of turns of the winch during lowering.  These two measurements were averaged together to determine the absolute pinger depth with $\pm$5~m uncertainty.  Due to a mistake made in converting winch turns to depth (later corrected and verified with the pressure sensors), the pinger was stopped at depths that are systematically shallower than the instrumented sensor depths, by an amount that increases with depth.  For the measurement with the sensor at 500~m depth, the pinger was at 477~m depth.  This has the effect that the relative uncertainty in the sound speed is $\pm$0.6\% for shallow depths (where it is dominated by the horizontal distance uncertainty) and increases to $\pm$1\% for deep depths (where the pinger depth uncertainty contributes nearly as much as the horizontal distance uncertainty).

While pressure and shear wave pulses were detected for many pinger-sensor combinations, for this analysis we selected only those with very high signal-to-noise ratio (SNR), sufficient to not only resolve the pulse but also to resolve its start time precisely.  For shear waves, only String D had sufficient SNR to identify the pulse start time precisely.  Within String D only the 5 non-HADES sensors had sufficient SNR, so there are 5 high-quality shear wave measurements.  For pressure waves, all 7 String D sensors, and 5 of 7 String B sensors, had runs with sufficient SNR.  This resulted in pressure wave measurements at 8 different depths, 4 of which have measurements with both strings.  For those depths with sound speed measured redundantly, the results agree well.

\subsection{Propagation time}
Each 9~s sensor waveform contains 9 pinger pulses, which were averaged together to increase the pulse SNR.  For each pulse sample both the mean and the standard deviation amplitude were determined.  This averaging procedure was designed to decrease the (incoherent) noise by a factor of 3 without affecting the signal amplitude.

While the pinger emission is driven by a clock which is continuously synchronized with GPS time, the sensor recording is driven by an analog-to-digital converter (ADC) clock which drifts by an amount on the order of 10 $\mu$s per second.  That is, the actual sampling frequency typically differs from the nominal sampling frequency by $\sim$10 parts per million.  Furthermore, the actual sampling frequency varies with time (the clock drift rate itself drifts).  This means that pulse averaging using the nominal time of each acquired sample results in large decoherence and a false average waveform.  This clock drift effect was removed by using the true absolute time of each sample as determined continuously from the IRIG-B GPS signal.  This is a 100 PPS digital signal that is sampled synchronously with the sensor voltage data.  Rising edges occur every 10~ms and pulse widths encode the absolute time.

After applying the clock drift correction algorithm, the absolute time of each sample of the waveform is known with a precision of $\pm$10~$\mu$s.  These absolute times were used in the pulse averaging: Absolute sample times were wrapped modulo the pulse repetition period (1 second), and were then sorted and binned to determine the average time and amplitude of each consecutive set of 9 samples.  Figure~\ref{waveform} shows a typical average waveform recorded by a sensor.

For each averaged waveform, a bipolar discriminator was applied to determine the start time.  The noise level varied too much from channel to channel to use a fixed threshold, but for each channel the first cycle of the pinger signal was clearly visible above the noise.  Therefore a threshold was manually chosen for each channel.  The first threshold crossing was then verified by eye to be a good estimate of the signal start time for each channel.  The uncertainty on this arrival time determination is estimated to be $\pm$0.05~ms, corresponding to $\sim$1 signal oscillation period.

The uncertainty of the emission time is simply that of the HV pulser time delay, including variation with temperature: $\pm$0.05~ms.

\section{Results}

\begin{figure}[tbp]
\centering
\noindent\includegraphics[width=20pc]{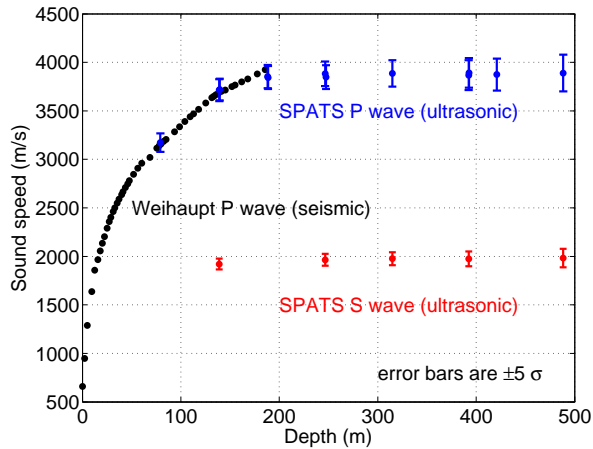}
\caption{Measurements of sound speed for both pressure and shear waves at particular depths using the South Pole Acoustic Test Setup featuring transmitters and sensors at $\sim$5-25~kHz.  A previous measurement made at seismic (Hz) frequencies~\cite{Weihaupt63} is shown for comparison.  Note: the SPATS error bars are $\pm$5~$\sigma$ in order to be visible.  No uncertainty estimate is available for the Weihaupt result.}
\label{spats_and_weihaupt}
\end{figure}

\subsection{Overview}

Figure~\ref{spats_and_weihaupt} shows our measurement of the sound speed versus depth for both pressure and shear waves.  A previous measurement of pressure wave speed in firn~\cite{Weihaupt63} is shown for comparison.  Table~\ref{error_budget} shows the error budget for two example data points in the analysis.

\subsection{Pressure waves}

\begin{figure*}
\begin{center}
\subfigure[][]{
\label{fits-a}
\noindent\includegraphics[width = 0.45 \textwidth]{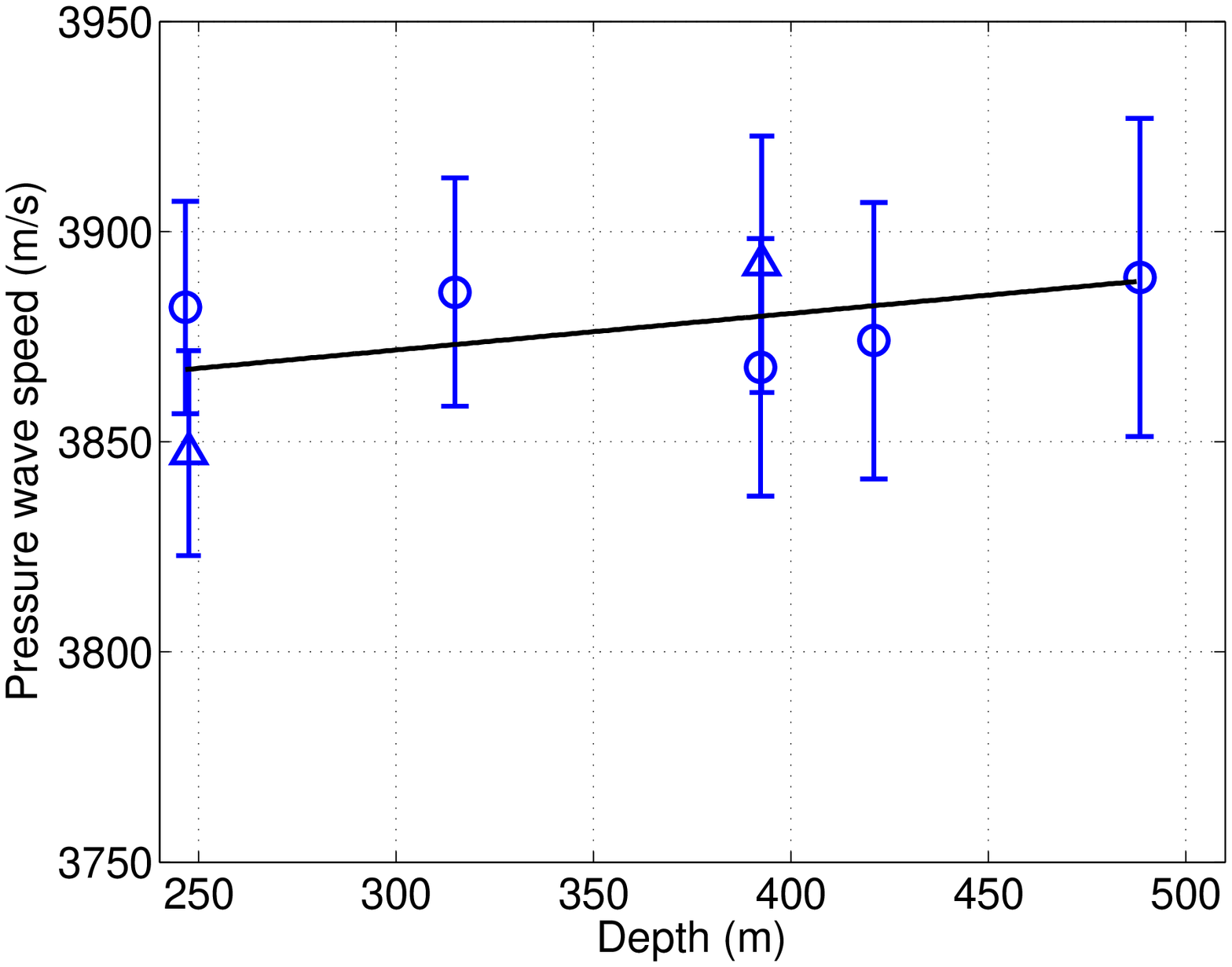}
}
\subfigure[][]{
\label{fits-b}
\noindent\includegraphics[width = 0.45 \textwidth]{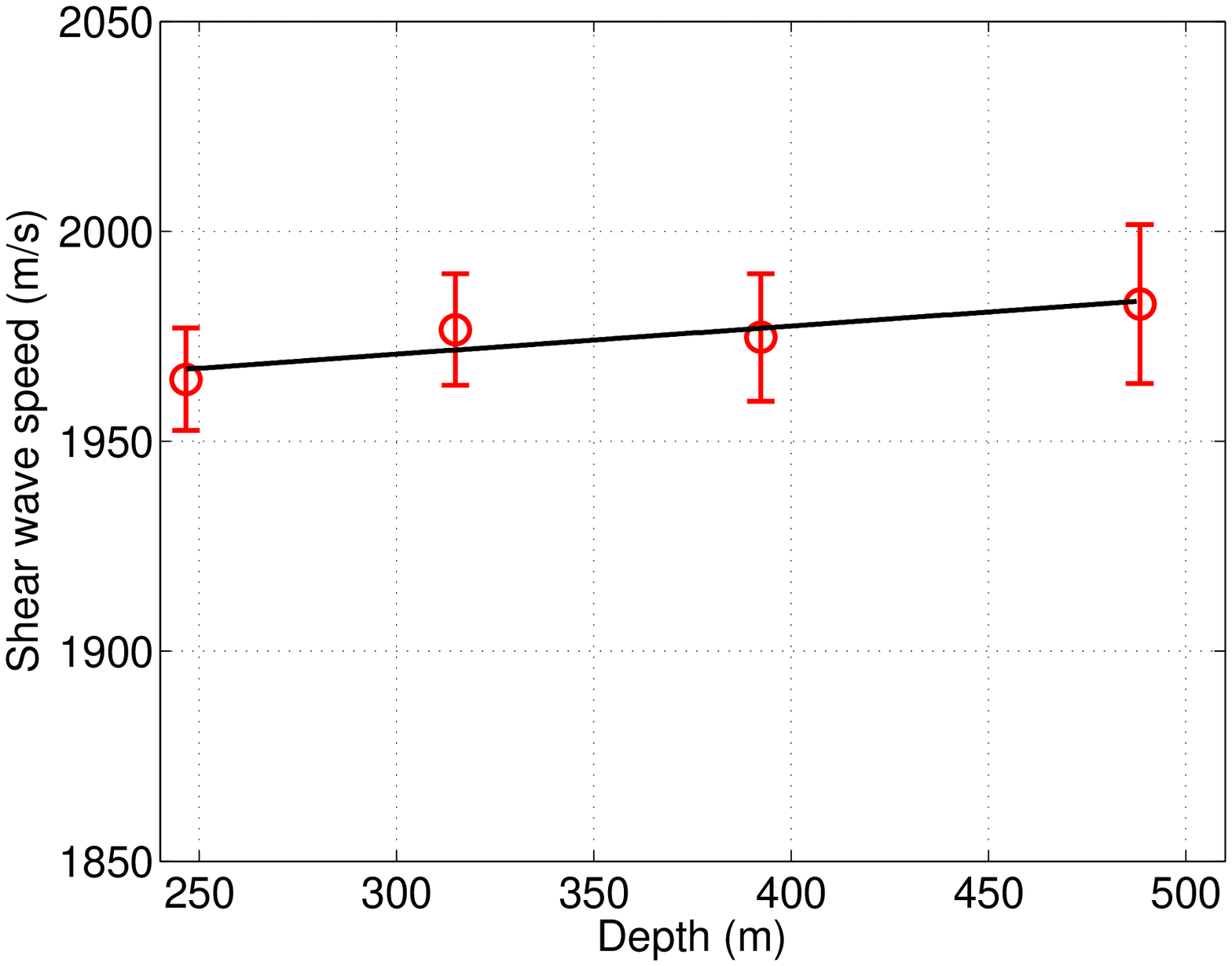}
}
\subfigure[][]{
\label{fits-c}
\noindent\includegraphics[width = 0.45 \textwidth]{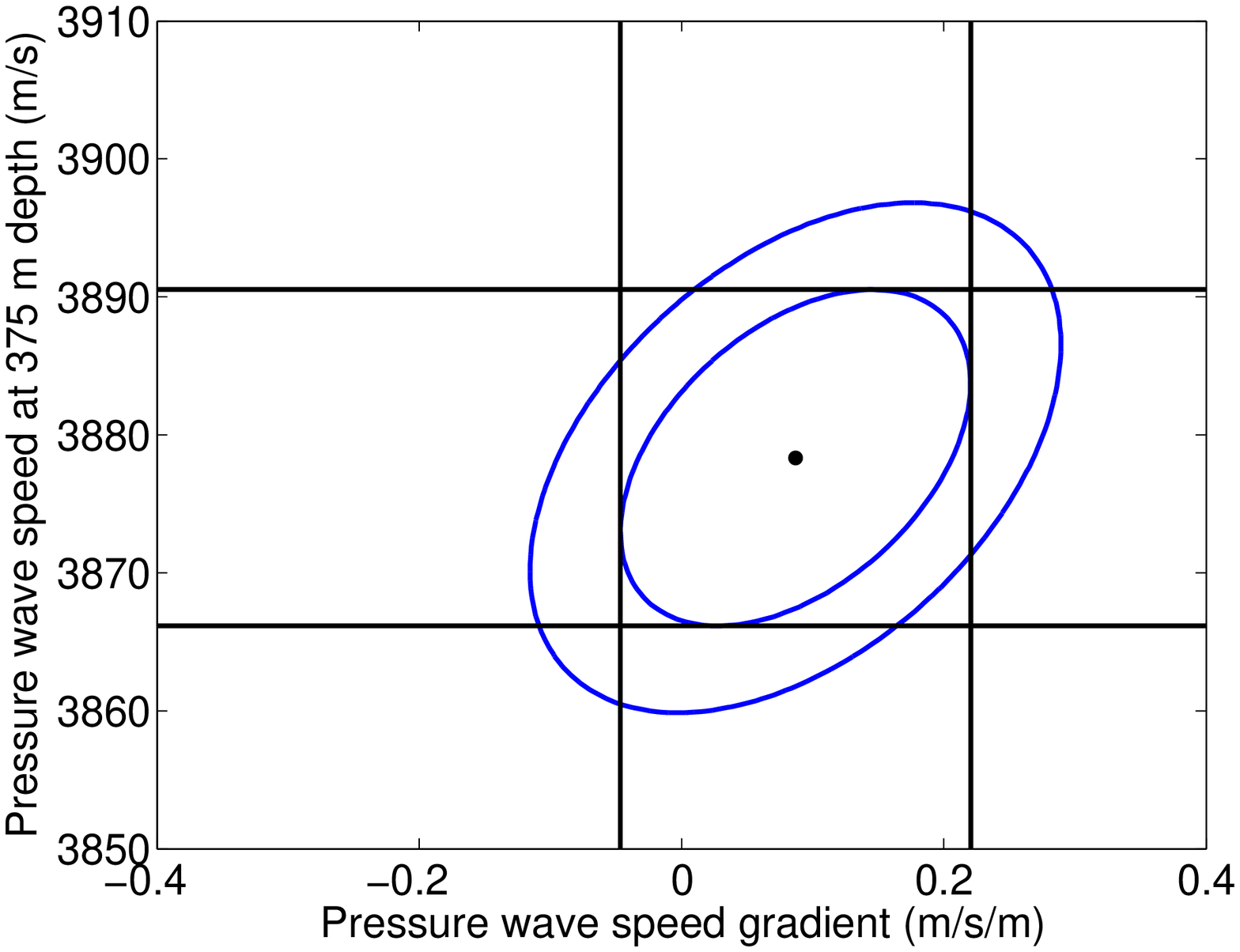}
}
\subfigure[][]{
\label{fits-d}
\noindent\includegraphics[width = 0.45 \textwidth]{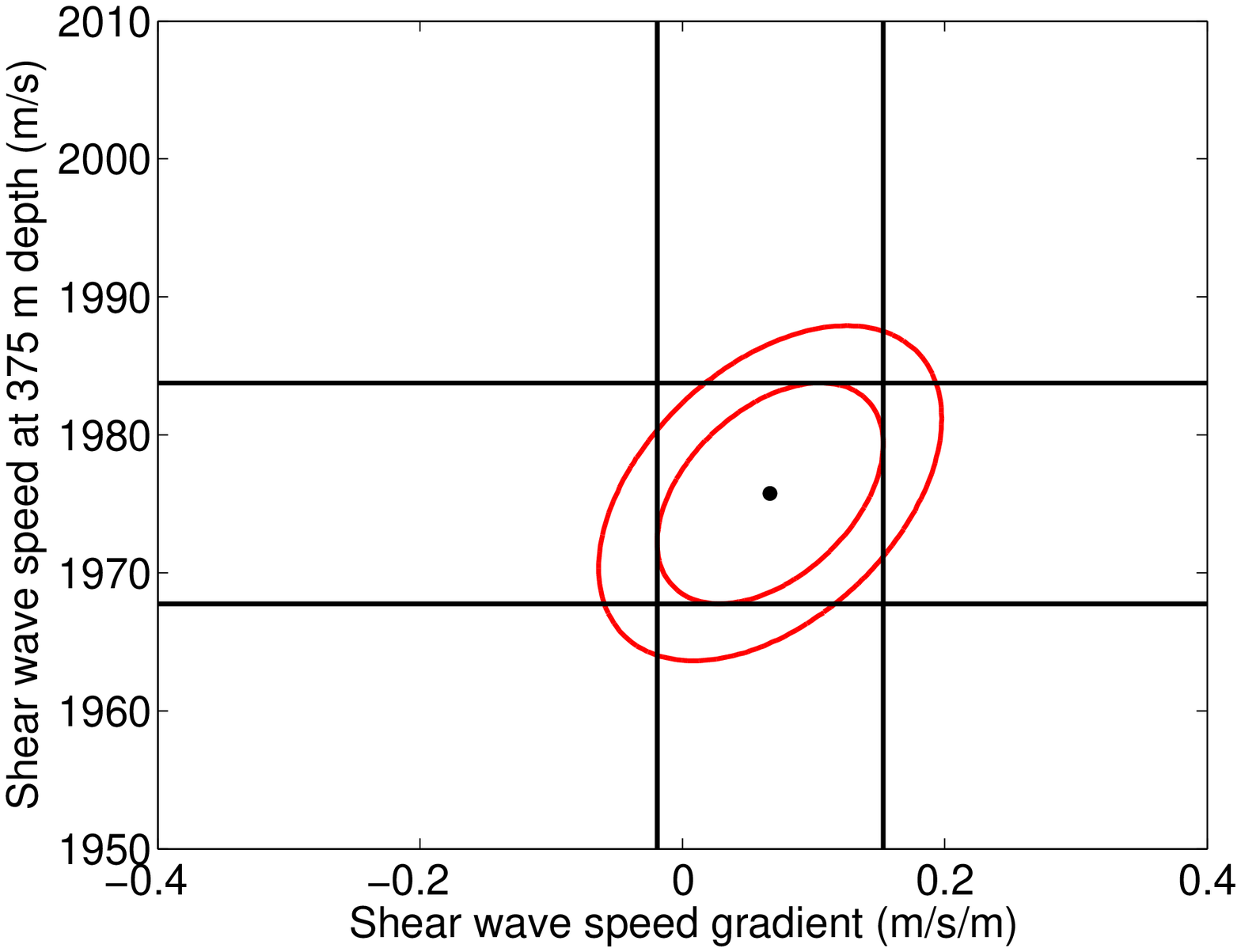}
}
\caption{Pressure wave \subref{fits-a} and shear wave \subref{fits-b} speed vs. depth between 250 and 500~m depth.  Error bars are $\pm$1~$\sigma$.  Measurements made with String D (B) are shown as circles (triangles).  Confidence regions for a joint fit of sound speed and sound speed gradient are also shown, for both pressure waves \subref{fits-c} and shear waves \subref{fits-d}.  The dot gives the best fit ($\chi^2$ = 1.61 for 7-2=5 degrees of freedom for P waves and $\chi^2$ = 0.195 for 4-2=2 degrees of freedom for S waves).  The inner (outer) contour is drawn for $\Delta\chi^2$ = 1.00 (2.30).  The outer ellipse encloses the most likely 68\% of parameter space for the two parameters fit jointly.  The horizontal (vertical) lines give the one-sigma confidence region for the sound speed (sound speed gradient) fit individually.  Note that all errors are treated as uncorrelated.  It is possible that the systematic error contributions from the pinger and sensor positions are correlated between different measurements and that this is why the $\chi^2$ values are smaller than expected for uncorrelated errors.}
\label{fits}
\end{center}
\end{figure*}

Figure~\ref{fits}\subref{fits-a} shows a close up of the pressure wave speed versus depth in the deep, fully compactified ice.  A linear fit was made to the data in the fully compactified region from 250 to 500~m depth:
\begin{linenomath*}
\begin{equation}
v_P(z) = [z - (375\textrm{ m})] \times g_P + v_P(375\textrm{ m}),
\end{equation}
\end{linenomath*}
where $z$ is the depth (measured positive downward from the surface), $v_P(z)$ is the pressure wave speed at depth $z$, and $g_P$ is the pressure wave speed gradient in the 250-500~m depth range.  The parameterization was chosen such that the sound speed in the center of the fitted range is one of the parameters.  The best fit is:
\begin{linenomath*}
\begin{equation}
v_P(375\textrm{ m}) = (3878 \pm 12) \textrm{ m/s};
\end{equation}
\end{linenomath*}
\begin{linenomath*}
\begin{equation}
g_P = (0.087 \pm 0.13) \textrm{ m/s/m}.
\end{equation}
\end{linenomath*}

Figure~\ref{fits}\subref{fits-c} shows our constraints on the two-parameter fit (sound speed and sound speed gradient) describing the pressure wave propagation as a function of depth in the fully compactified (bulk) ice.  The gradient is consistent with zero.

In the firn, our pressure speed results are consistent with the previous measurements by Weihaupt.

\subsection{Shear waves}

Figure~\ref{fits}\subref{fits-b} shows a close up of the shear wave speed versus depth in the deep, fully compactified ice.  A linear fit was performed to the data in the fully compactified region from 250 to 500~m depth:
\begin{linenomath*}
\begin{equation}
v_S(z) = [z - (375\textrm{ m})] \times g_S + v_S(375\textrm{ m}),
\end{equation}
\end{linenomath*}
where $v_S(z)$ is the shear wave speed at depth $z$, and $g_S$ is the shear wave speed gradient.  The best fit is:
\begin{linenomath*}
\begin{equation}
v_S(375\textrm{ m}) = (1975.8 \pm 8.0) \textrm{ m/s};
\end{equation}
\end{linenomath*}
\begin{equation}
g_S = (0.067 \pm 0.086) \textrm{ m/s/m}.
\end{equation}

Figure~\ref{fits}\subref{fits-d} shows our constraints on the two-parameter fit (sound speed and sound speed gradient) describing the shear wave propagation as a function of depth in the fully compactified (bulk) ice.  The gradient is consistent with zero.

The shallowest depth for which we have a precise shear wave determination is 139~m depth.  At this depth the ice is still not fully compactified.  As expected, the shear wave speed at this depth (1921~$\pm$~11~m/s) is slower than in the deep ice.
 
\section{Measurement with explosives}

In addition to the precision measurement using piezoelectric transmitters and sensors at $\sim$5-25~kHz, a complementary measurement was performed with explosives (seismic frequencies).  This measurement was performed in January 1999 as part of deployment of the AMANDA neutrino telescope.  Dynamite was attached to detonation cord and lowered to a depth $z_d = $~50~$\pm$~5~m in a mechanically drilled hole.  This hole was located $\sim$15~m horizontally from AMANDA Hole 13, which had an acoustic sensor (hydrophone) at depth $z_h = $~2250~$\pm$~10~m.  The detonation cord had an active core of PETN (pentaerythritol tetranitrate).  An electrical circuit near the blasting cap end of the cord triggered a digital oscilloscope to start recording the hydrophone signal.

A pulse was clearly visible above the noise, at an arrival time $t_a =$~566~$\pm$~5~ms after the trigger.  Assuming the detonation signal propagated\footnote{The detonation cord was ``Red Cord'' from Imperial Chemical Industries (ICI).  ICI was purchased by Orica Mining Services Worldwide, which now makes the same cord under the name ``Cordtex Pyrocord Detonating Cord.''  The detonation velocity we use is taken from the manufacturer's data sheet.} at $v_d =$~6750~$\pm$~250~m/s through the cord of length $L = $~52~$\pm$~3~m, the measured pressure wave speed is
\begin{linenomath*}
\begin{equation}
v_P = \frac{z_h-z_d}{t_a-L/v_d} = 3941 \pm 41 \textrm{ m/s.}
\end{equation}
\end{linenomath*}

The precision achieved in this measurement is $\pm$1.0\%.  It gives the pressure wave speed averaged over the depth profile from 50 to 2250~m depth.  It is consistent with the result obtained from the piezeoelectric instrumentation, despite the significantly different frequency band and depth range.  This integral, vertical measurement is complementary to the differential, horizontal measurement.  It provides a valuable cross check and extends the range of measurement to nearly the entire thickness (2.8~km) of the South Pole ice.  The explosives measurement indicates that the pressure wave speed gradient is small not only in the 200-500~m depth range but also down to $\sim$2~km depth.

\begin{table}[tbp]	
\centering
\caption{Error budget for two example data points.  Each is for pinging from Hole 69 and receiving on String D.  While the error contribution from the horizontal distance is nearly the same for all data points, the contribution from the pinger depth increases with depth.  This is because the vertical distance between the pinger and sensor increases with depth, and the error contribution is proportional to this difference.} 
\centering      
\begin{tabular}{|  c  |  c  |  c  |}  
\hline                        
 & \bf{P wave} & \bf{S wave} \\
\hline
sensor depth (m) & 500 & 140 \\
\hline
pinger depth (m) & 477 & 138 \\
\hline
sound speed (m/s) & 3889 & 1921 \\
\hline
error due to horizontal distance (\%) & 0.54 & 0.57 \\
\hline
error due to pinger depth (\%) & 0.72 & 0.06 \\
\hline
error due to sensor depth (\%) & 0.29 & 0.02 \\
\hline
error due to emission time (\%) & 0.15 & 0.08 \\
\hline
error due to arrival time (\%) & 0.15 & 0.08 \\
\hline
\bf{total error} (\%) & \bf{0.97} & \bf{0.58} \\
\hline
\end{tabular} 
\label{error_budget}
\end{table} 

\section{Refraction}

\subsection{Calculated ray trajectories in firn and bulk ice}
\begin{figure}[tbp]
\centering
\noindent\includegraphics[width=20pc]{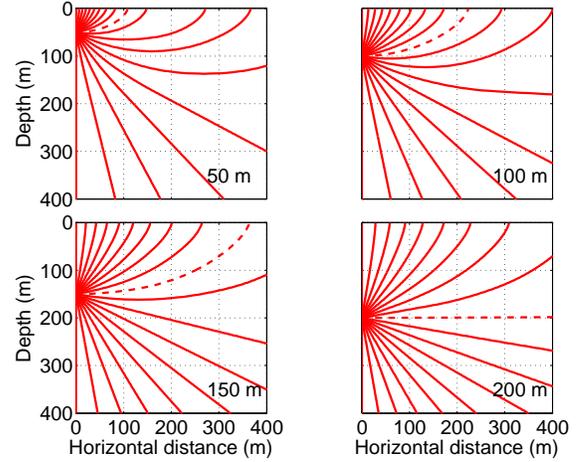}
\caption{Calculated pressure wave ray trajectories using the measured sound speed as a function of depth.  Refraction is significant in the firn (shallower than $\sim$174~m) and negligible below it.  Each panel shows rays emitted from a source at 50, 100, 150, or 200~m.  Rays are emitted every 10$^\circ$ from vertical upward to vertical downward.  The horizontally emitted ray is indicated by a dashed line.  The Weihaupt profile is used for depths between 0 and 174~m, and the SPATS linear best fit is used for depths between 174 and 500~m.  Although the two results agree within their error bars in the region from 174-186~m depth, the SPATS best fit predicts a sound speed slightly smaller than the Weihaupt results.  The two curves intersect at 174~m and the SPATS fit is chosen in the overlap region so that there is no kink in the velocity profile.}
\label{ray_traces}
\end{figure}

We have calculated the trajectory of individual acoustic rays to illustrate the degree of refraction for various source depths and emission directions.  Figure~\ref{ray_traces} shows example ray trajectories calculated for pressure waves.  The ray tracing was performed using an algorithm~\cite{Boyles84} that treats the ice as a layered medium, in each layer of which the sound speed gradient is constant.  Because the gradient is constant in each thin layer, the ray segment in each layer is an arc of a circle.  This algorithm gives a fast and accurate piecewise second order approximation to the true ray path and simultaneously calculates the integrated path length and travel time.  Note that in the presence of a vertical velocity gradient, even horizontally emitted rays are refracted toward the direction of decreasing sound speed.

\subsection{Radius of curvature in bulk ice}
Because the trajectory of a ray in a medium with constant sound speed gradient is a circle, a convenient way to quantify the amount of refraction is the radius of curvature:
\begin{linenomath*}
\begin{equation}
R = \frac{v}{|g|},
\end{equation}
\end{linenomath*}
where $v$ is the sound speed and $g$ is the gradient of the sound speed.  For pressure (shear) waves, our joint fit for the sound speed and sound speed gradient gives a best fit radius of curvature of $\sim$44~km ($\sim$29~km).  With a 44~km radius of curvature, a ray of length $L$=1~km (a possible propagation distance from source to sensor in a large neutrino detector) deflects by a small amount d with respect to straight-line propagation:
\begin{linenomath*}
\begin{equation}
d=\frac{L^2}{2R} = 11\textrm{ m}.
\end{equation}
\end{linenomath*}
This amount of deflection is smaller than the thickness of the radiation pattern induced by a neutrino.  Note that this is the deflection predicted using our best fit gradient.  Because our measurement of the gradient is consistent with zero, the radius of curvature is also consistent with infinity (zero deflection).

\section{Discussion}

\subsection{Comparison with previous results}

Kohnen~\cite{Kohnen} compiled sound speed measurements from Antarctica and Greenland.  After applying quality selection criteria to the existing measurements, he found a simple dependence of both pressure wave speed and shear wave speed on temperature: $v_p = -(2.30 \pm 0.17) T + 3795$ and $v_s = -(1.2 \pm 0.58) T + 1915$, where $v_p$ is the pressure wave speed in m/s, $v_s$ is the shear wave speed in m/s, and T is the temperature in~$^\circ$C. 

\begin{figure}[tbp]
\centering
\noindent\includegraphics[width=20pc]{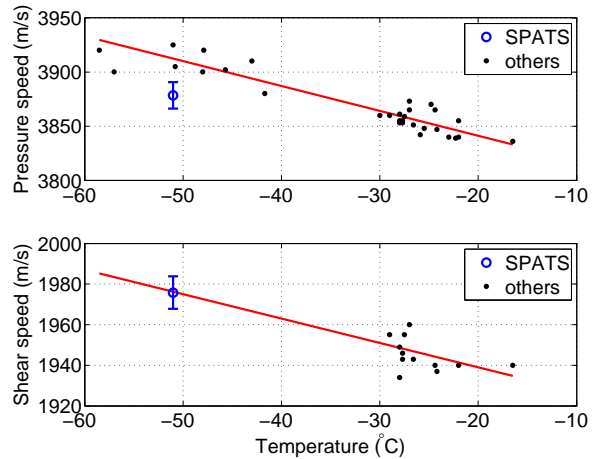}
\caption{Compilation of sound speed vs. temperature in ice from different authors.  Only field (not laboratory) measurements are shown, and only measurements in the Greenland and Antarctic ice sheets (not temperate glaciers) are shown.  The non-SPATS data compilation is taken from~\cite{Kohnen}.  Previously the only shear wave measurements were between -15 and -30~$^\circ$C; SPATS has extended this to -51~$^\circ$C.  The lines give Kohnen's fits, without re-fitting to include SPATS.  The SPATS pressure wave result is slightly slower than previous results.  The shear wave result matches the low-temperature prediction of Kohnen (made by assuming Poisson's ratio is temperature independent, predicting a shear wave speed corresponding to each pressure wave speed, and fitting a straight line) very well.}
\label{speed_vs_temp}
\end{figure}

Figure~\ref{speed_vs_temp} shows the data points compiled by Kohnen along with the new SPATS measurement reported here.  Our pressure wave speed is slightly slower than the other measurements.  The other measurements do not include error estimates, so it is difficult to determine whether our result is consistent with them.  The other measurements were made with refraction shooting, in which rays are traced from a surface explosion to a surface sensor, and the maximum speed below the firn is deduced by unfolding the refraction through the firn.  Our \emph{in situ} measurement is less susceptible to systematic effects because it uses unrefracted rays between sources and sensors buried in the deep fully compactified ice.

The SPATS shear wave measurement is the first below -30~$^\circ$C.  The shear wave fit by Kohnen was made using predictions at low temperature from the pressure speed and assuming temperature-independent Poisson's ratio.  Our measurement agrees well with his prediction.

A laboratory measurement of both pressure and shear wave speed in ice was reported recently~\cite{Vogt08}.  A degassing system was used to produce a $\sim$3~m$^3$ block of bubble-free ice in which the speeds were measured between 0 and -20~$^\circ$C.  The measured speeds were larger than predicted from the Kohnen fit by $\sim$50~m/s, perhaps due to the absence of bubbles or to grain orientation in the laboratory measurement.

The SPATS and Weihaupt results for pressure wave speed in South Pole firn are consistent in their region of overlap.  This is a valuable cross check because the two measurements use very different experimental methods and use signal frequencies that differ by 4 orders of magnitude.

While Weihaupt measured the pressure wave speed to a maximum of 186~m depth, all his measurements were in the firn ice (by necessity, because his measurement used waves that were refracted back to the surface).  We have confirmed Weihaupt's measurement in the firn and extended it into the fully compactified bulk ice, to a maximum depth of 500~m.  Moreover, we have for the first time measured the shear wave speed in South Pole ice and have done so both in the firn and bulk ice, at depths from 140 to 500~m depth.

\subsection{Implications for neutrino astronomy}

We have determined that the sound speed gradient in deep South Pole ice is consistent with zero, and therefore that the amount of refraction of acoustic waves is consistent with zero.  This is in contrast with most deep ocean sites, where refraction due to a vertical sound speed gradient is a significant challenge for acoustic neutrino detection~\cite{Vandenbroucke05}.

Optical photons are scattered in the ice such that typical photons detected in the AMANDA and IceCube arrays have scattered several times, losing much of their directionality.  Sophisticated algorithms have been developed to reconstruct the neutrino direction and energy, and the interaction location, in the presence of scattering~\cite{Ahrens04}.  Typically these algorithms fit the full scattered waveform shape and then use the rising edge to determine the arrival time of the ``direct'' unscattered photons.

Radio waves are refracted significantly in the firn and negligibly in the deep ice.  The RICE experiment spans both the firn and the bulk ice and therefore must account for refraction in its signal reconstruction, background rejection, and neutrino sensitivity determination.  We note that while radio waves are refracted downward in the firn, acoustic waves are refracted upward.  This means that while surface radio noise is waveguided down to a possible deep detector, surface acoustic noise is refracted back to the surface, such that the firn shields deep sensors from surface noise.  This expectation is confirmed by the observation that SPATS sensor ambient noise levels vary negligibly with time, despite the operation of large construction equipment directly above the array during each South Pole season~\cite{Karg09}.

We have shown that acoustic waves, similar to radio waves, propagate unrefracted in deep South Pole ice.  This means the location of an acoustic source can be reconstructed quickly and precisely using analytical methods.  Furthermore, our measurements imply that the acoustic radiation pattern (like the radio radiation pattern) is affected negligibly by refraction.  This unique pattern (a wide, flat ``pancake'') could be used as a signature to separate neutrino events from background events, which are likely to produce a spherical radiation pattern.  The radiation pattern could also be used (along with signal arrival time and amplitude) for neutrino event reconstruction.  For example, the neutrino arrival direction could be estimated by fitting a plane to the hit sensors; the upward normal points to the neutrino source~\cite{Vandenbroucke06Hybrid}.

We note that a similar array to RICE, deployed beneath the firn to avoid refraction, would benefit similarly to the acoustic method from preserved radiation pattern.  In fact, codeploying acoustic and radio arrays in the same volume of ice could allow the two arrays to operate in hybrid mode, detecting a significant fraction of events in coincidence~\cite{Besson05}.

If both a pressure and a shear wave pulse from a neutrino are detected by a single acoustic sensor, the time difference between them could be used to estimate the distance to the source and from this the neutrino energy, with a single sensor.  For distances less than $\sim$100~m, the precision of this reconstruction is dominated by the pulse arrival time resolution.  If the timing resolution is $\sim$0.1~ms, the distance resolution is $\sim$1~m, independent of distance within this ``near'' regime.  For distances larger than $\sim$100~m, the distance reconstruction precision is limited by the precision of our sound speed measurement.  Using the $\sim$1\% sound speed measurement presented here, the distance could be determined with $\sim$1\% precision (that is, the distance precision scales with distance in this ``far'' regime).

Now that the sound speed profile has been determined, it remains to determine the attenuation length and absolute noise level of South Pole ice, to determine its potential for acoustic neutrino detection.  Data taking is ongoing with the SPATS array to achieve this goal.

SPATS is now operating in a mode to trigger on ambient impulsive transients.  Such transients have been detected in coincidence between multiple sensors hundreds of meters apart.  Using the sound speed profile presented here, the location and emission time of these acoustic sources can be reconstructed precisely~\cite{Vandenbroucke08}.



\section{Acknowledgments}
We acknowledge the support from the following agencies:
U.S. National Science Foundation-Office of Polar Program,
U.S. National Science Foundation-Physics Division,
University of Wisconsin Alumni Research Foundation,
U.S. Department of Energy, and National Energy Research Scientific Computing Center,
the Louisiana Optical Network Initiative (LONI) grid computing resources;
Swedish Research Council,
Swedish Polar Research Secretariat,
and Knut and Alice Wallenberg Foundation, Sweden;
German Ministry for Education and Research (BMBF),
Deutsche Forschungsgemeinschaft (DFG), Germany;
Fund for Scientific Research (FNRS-FWO),
Flanders Institute to encourage scientific and technological research in industry (IWT),
Belgian Federal Science Policy Office (Belspo);
the Netherlands Organisation for Scientific Research (NWO);
Marsden Fund, New Zealand;
M.~Ribordy acknowledges the support of the SNF (Switzerland);
A.~Kappes and A.~Gro{\ss} acknowledge support by the EU Marie Curie OIF Program;
J.~P.~Rodrigues acknowledge support by the Capes Foundation, Ministry of Education of Brazil.



\bibliography{spats_sound_speed}
\bibliographystyle{elsart-num}


\end{document}